\documentclass[11pt]{article}
\usepackage{amsmath}
\usepackage{amssymb}

\newtheorem{theorem}{Theorem}[section]

\newtheorem{define}[theorem]{Definition}

\newsavebox{\savepar}
\newenvironment{boxit}{\begin{center} \begin{lrbox}{\savepar}
        \begin{minipage}[]{\textwidth}}
        {\end{minipage}\end{lrbox}
      \fbox{\usebox{\savepar}
} \end{center}}

\begin{document}

\title{A Provably Secure Ring Signature Scheme in\\ Certificateless Cryptography
\thanks{\quad The main part of this work is published in \cite{ZZW07}.}}

\author{\textsc{Lei Zhang $^a$
\thanks{\quad Email: leizhang@sei.ecnu.edu.cn}
\quad  Futai Zhang $^b$
%\thanks{\quad zhangfutai@njnu.edu.cn}
\quad Wei
Wu $^c$
%\thanks{\quad weiwu81@gmail.com}
}
\\
\small{$^a$ East China Normal
University, Shanghai, China}\\
\small{$^b$ Nanjing Normal University, Nanjing, China}\\
\small{$^c$ Fujian Normal University, Fuzhou, China}}
\date{}
\maketitle

%%%%%%%%%%%%%%%%%%%%%%%%%%%%%%%%%%%%%%%%%%%%%%%%%%%%%%%%%%%%%%%%%%%%%%%%%%%%%
%%%%%%  We need a short abstract.
%%%%%%%%%%%%%%%%%%%%%%%%%%%%%%%%%%%%%%%%%%%%%%%%%%%%%%%%%%%%%%%%%%%%%%%%%%%%%
\begin{abstract}
Ring signature is a kind of group-oriented signature. It allows a
member of a group to sign messages on behalf of the group without
revealing his/her identity. Certificateless public key cryptography
was first introduced by Al-Riyami and Paterson in Asiacrypt 2003. In
certificateless cryptography, it does not require the use of
certificates to guarantee the authenticity of users' public keys.
Meanwhile, certificateless cryptography does not have the key escrow
problem, which seems to be inherent in the Identity-based
cryptography. In this paper, we propose a
concrete certificateless ring signature scheme. The security models
of certificateless ring signature are also formalized. Our new
scheme is provably secure in the random oracle model, with the
assumption that the Computational Diffie-Hellman problem is hard. In
addition, we also show that a generic construction of
certificateless ring signature is insecure against the key
replacement attack defined in our security models.
\end{abstract}

%%%%%%%%%%%%%%%%%%%%%%%%%%%%%%%%%%%%%%%%%%%%%%%%%%%%%%%%%%%%%%%%%%%%%%%%%%%%%
%%%%%%%%%%%%%%%%%%%%%%%%%%%%%%%%%%%%%%%%%%%%%%%%%%%%%%%%%%%%%%%%%%%%%%%%%%%%%
%%%%%%
%%%%%%  From here the main part of your paper begins
%%%%%%
%%%%%%%%%%%%%%%%%%%%%%%%%%%%%%%%%%%%%%%%%%%%%%%%%%%%%%%%%%%%%%%%%%%%%%%%%%%%%
%%%%%%%%%%%%%%%%%%%%%%%%%%%%%%%%%%%%%%%%%%%%%%%%%%%%%%%%%%%%%%%%%%%%%%%%%%%%%

\section{Introduction}
\label{sec-0}

In Asiacrypt 2001, Rivest, Shamir and Tauman \cite{R. Rivest}
introduced the concept of ring signature, which makes it possible to
specify a set of possible signers without revealing which member
actually produced the signature. As pointed in \cite{R. Rivest},
ring signatures provide an elegant way to leak authoritative secrets
in an anonymous way, to sign casual email in a special way that can
only be verified by its intended recipient, anonymous membership
authentication for ad hoc groups \cite{Bresson}, etc. In addition,
ring signatures can also be served as the building block of
concurrent signatures and solve some other problems in multiparty
computations.

Ring signatures can be regarded as the simplified group signatures
that have only users and no managers. Group signatures are useful in
the situation where the members want to cooperate, while ring
signatures are useful when the members do not want to cooperate.
Both group signatures and ring signatures are
\emph{signer-ambiguous}. However, group signatures have the
additional feature that the anonymity of a signer can be revoked
(i.e., the signer can be traced) by a designated group manager. Ring
signatures allow greater flexibility: no centralized group manager
or coordination among the various users is required (indeed, users
may be unaware of each other at the time they generate their public
keys); rings may be formed completely in an ad-hoc manner, do not
require any coordination among the various users (indeed, users do
not even need to be aware of each other) and full anonymity (unless
the actual signer decides to expose himself). To produce a ring
signature, the actual signer declares an arbitrary set of possible
signers that must include himself, and computes the signature
entirely by himself using only his private key and the others'
public keys.\\
%%%%%%%%%%%%%%%%%%%%%%%%%%%%%%%%%%%%%%%%%%%%%%%%%%%%%%%%%%%%%
%                                                           %
%                     Traditional PKC                       %
%                                                           %
%%%%%%%%%%%%%%%%%%%%%%%%%%%%%%%%%%%%%%%%%%%%%%%%%%%%%%%%%%%%%
\indent In traditional public key cryptosystem (PKC), each user $U$
has a pair of cryptographic keys--a public key and a private key.
The private key is kept secret by the user himself, while the public
key may be widely distributed. Anyone can encrypt messages with
$U$'s public key and obtain the ciphertexts  which can only be
decrypted with $U$'s private key. Similarly, one can use $U$'s
public key to verify if a signature is generated by $U$. Therefore,
there is no need for the sender and receiver to share secret
information before the communication. The biggest challenge in PKC
is ensuring the authenticity of public key, that is how to bind  a
user and his public key. Suppose Alice wants to encrypt a message to
send to Bob, and Bob is someone who Alice does not know personally,
how can Alice be sure that Bob's purported public key really is
Bob's key (and not Charlie's, for example). If Alice uses a ``false"
public key to encrypt the message and send the ciphertext to Bob, it
will result that the intended receiver Bob can not obtain the
message, and even worse, someone else can decrypt the ciphertext and
read the message. The usual approach to solve this problem is to use
a public key infrastructure (PKI), in which one or more third
parties, known as certificate authorities, issue certificates to
bind a user and his public key. In traditional PKC, one must first
check the authenticity of the pair $(U,PK)$ by verifying the
validity of its certificate before any operation regarding the user
$U$. History has shown that certificates in traditional PKC are
generally considered to be costly to use and manage. It is even more
problematic for a ring signature scheme in traditional PKC,  where
the signer must first verify all the certificates of group members
before generating the ring signature on behalf of that group,
otherwise his anonymity is jeopardized under the extreme case that
all other ring members' certificates  are indeed invalid. Given a
ring signature, the verifier must perform the same verification as
well before checking the validity of the ring signature. This will
lead to the inefficiency of the whole scheme because the computation
cost will
increase linearly with the group size.\\
%%%%%%%%%%%%%%%%%%%%%%%%%%%%%%%%%%%%%%%%%%%%%%%%%%%%%%%%%%%%%
%                                                           %
%                  Identity-based PKC                       %
%                                                           %
%%%%%%%%%%%%%%%%%%%%%%%%%%%%%%%%%%%%%%%%%%%%%%%%%%%%%%%%%%%%%
\indent In 1984, Shamir \cite{Shamir} first proposed the
Identity-Based public key cryptography (ID-PKC), in which the public
key of a user is some unique public information about the identity
of the user (e.g. a user's email address) \cite{DLZZS17,LZ13}. Therefore, the need of
certification can be eliminated. A Trusted Third Party, called the
Private Key Generator (PKG), generates the corresponding private
keys for the users in ID-PKC. To operate,  the PKG first publishes a
``master" public key, and keeps the corresponding master private key
as secret. Given the master public key, any party can compute a
public key corresponding to an identity $ID$ by combining the master
public key with the identity value. To obtain a corresponding
private key, the party authorized to use the identity $ID$ contacts
the PKG, which uses the master private key to generate the private
key for the identity $ID$.  However, this approach creates a new
inherent problem, namely the {\em key escrow} of a user's private
key, since PKG must be {\em completely trusted}. This is due to the
knowledge of the PKG on the user's private key. For a ring signature
scheme in ID-PKC, a malicious PKG can forge a ring
signature on behalf of any group without being detected.\\
%%%%%%%%%%%%%%%%%%%%%%%%%%%%%%%%%%%%%%%%%%%%%%%%%%%%%%%%%%%%%
%                                                           %
%                 Certificateless PKC                       %
%                                                           %
%%%%%%%%%%%%%%%%%%%%%%%%%%%%%%%%%%%%%%%%%%%%%%%%%%%%%%%%%%%%%
\indent In order to enjoy the implicit certification property of
ID-PKC while without suffering from its inherent key escrow problem,
Al-Riyami and Paterson \cite{Al-Riyami} proposed a new paradigm
called certificateless public key cryptography (CL-PKC). Different
from ID-PKC, a third party which we call Key Generation Center (KGC)
in CL-PKC does not have the access to a user's private key. Instead,
the KGC supplies a user with a partial private key, which derives
from the user's identity. Then the user combines the partial private
key with some secret information chosen by himself to generate his
actual private key. The corresponding public key is computed from
the system's public parameters and the secret information chosen by
the user, which is finally published in the system. Hence, it is no
longer an identity-based cryptography, since the public key needs to
be provided (but in contrast to the traditional public key
cryptography, the public key does not require any certificate).

Due to the lack of certification in CL-PKC, it is conceivable that
the adversary can replace anyone's public key of his choice. This
key replacement attack is also called {\em Type I adversary} in
\cite{Al-Riyami}. Obviously, a secure signature signature scheme in
CL-PKC must has the property that it is infeasible for {\em Type I
adversary} to create a valid signature under the false public key
chosen by the adversary himself. An assumption that must be made is
that KGC does not mount a public key replacement attack to a target
user since he is armed with this user's partial private key.
However, KGC might engage in other adversarial activities:
eavesdropping on signatures and making signing queries, which is
also known as {\em Type II Adversary}. In this way, the level of
trust is similar to the trust in a CA in a traditional PKI.

\subsection{Motivations} Certificateless cryptography have some
advantages over traditional PKC and ID-PKC in some aspects \cite{agg1,agg2}. As a
useful primitive, ring signatures have been studied in traditional
PKC and ID-PKC for more than five years. Even in a theoretic point
of view, ring signatures should be studied in CL-PKC to rich the
theories and techniques of CL-PKC. In practice, to generate a ring
signature on behalf of a group in traditional PKC, the signer must
first verify all the certificates of the group members, otherwise
his anonymity is jeopardized and the ring signature will be rejected
if he uses invalid certificates of some group members. Given a ring
signature, the verifier must perform the same verification as well
before checking the validity of the ring signature. These
verifications inevitably lead to the inefficiency of the whole
scheme since the computational cost increases linearly with the
group size. Although Identity-based ring signatures eliminate such
costly verifications, they suffer from a security drawback induced
by the inherent key escrow problem of ID-PKC. Namely, a malicious
PKG can always issue valid ring signatures on behalf of any group.
As CL-PKC does not use public key certificates, and in the meantime,
it removes the key escrow problem of ID-PKC, we think it supplies an
appropriate environment for implementing ring signatures. So it is
necessary to extend the notion and security model of ring signatures
to CL-PKC. Compared with ring signature schemes in traditional PKC,
in a CL-Ring scheme, both the signer and the verifier can avoid the
costly verification of group members' certificates. On the other
hand, in contrast to ID-based ring signatures, the KGC can no longer
forge a ring signature on behalf of a group without being detected.

In application aspects, like ring signatures in  traditional PKC and
ID-PKC, certificateless ring signatures can also be used in leaking
authoritative secrets in an anonymous way, anonymous membership
authentication for ad hoc groups \cite{Bresson}, reports to the
authorities embezzlement and corruption, certificateless designated
signatures and concurrent signatures, etc.

\subsection{Our Contributions} In this paper, we introduce the notion of ring
signature into certificateless cryptography and propose a concrete
certificateless ring signature scheme.

Firstly, we provide the security models of certificateless ring
signatures. Two types of adversaries: Type I adversary
$\mathcal{A}_{I}$ and Type II adversary $\mathcal{A}_{II}$ have been
formally defined. The above two adversaries in our definition are
``super adversaries" \cite{ZZ08}. That is, the
adversary can get valid ring signatures of the group whose public
keys have been replaced, without supplying the secret values that
are used to generate those public keys. In addition, our models also
capture the group-changing attack \cite{J. Liu} in the notion of
ring signatures.

Secondly, we give an analysis of a ``seem-secure" generic
construction of certificateless ring signatures. The generic
construction of certificateless signatures was first proposed by Yum
and Lee \cite{Yum}, which has been shown insecure in \cite{Hu}. Hu
{\em et al.} also presented a secure construction of certificateless
signatures \cite{Hu}. Using the similar methods in \cite{Hu}, one
can also get a generic construction of certificateless ring
signatures. However, as we will show later, the resulting generic
construction of certificateless ring signatures is totally insecure
against the key replacement attack.

Lastly, we present a concrete construction of certificateless ring
signatures. The new scheme uses the bilinear pairing on elliptic
curves and concretely, the signing phase requires 2 pairings and the
verification requires 3 pairings. We prove its security in the
random oracles, with the assumption that Computational
Diffie-Hellman problem is intractable.

\medskip
\noindent \textbf{Organization}.\\ The rest of the paper is
organized as follows. In the next section, we review some
preliminaries which are required in this paper.
Section~\ref{Sub-Adver-Model} defines the security models in the
notion of certificateless ring signatures. We analyze a generic
construction of certificateless ring signatures and show its
insecurity in Section~\ref{Sec-Analysis-Generic-Construction}. The
concrete construction of certificateless ring signature is proposed
in Section~\ref{Sec-Scheme}. Its security proofs are given in
Section~\ref{Sec-Proof}. Finally, Section~\ref{Sec-Conclusion} comes
our conclusion.

\section{Related Work}
%%%%%%%%%%%%%%%%%%%%%%%%%%%%%%%%%%%%%%%%%%%%%%%%%%%
%                                                 %
%                 Ring Signatures                 %
%                                                 %
%%%%%%%%%%%%%%%%%%%%%%%%%%%%%%%%%%%%%%%%%%%%%%%%%%%
Following the prior work of Rivest, Shamir and Tauman \cite{R.
Rivest}, a number of constructions of ring signature in traditional
PKC and ID-PKC have been presented. Abe, Ohkubo, and Suzuki \cite{M.
Abe} provided a construction applicable for several categories of
public keys (e.g., integer factoring based and discrete-log based).
A simple ring signature using bilinear maps was given in \cite{Dan
Boneh}. Herranz and Saez \cite{Herranz F} generalized the forking
lemma to the ring signatures. In \cite{Zhang Kim}, Zhang and Kim
extended the concept to Identity-Based ring signature (IDRS)
schemes. 
%Later Herranz and Saez \cite{Herranz}, Chow et al.
%\cite{S.M. Chow}, Chow and D. Wong \cite{chow 4} presented some
%efficient IDRS schemes respectively. In \cite{S.M. Chow2} Chow el
%al. gave a solid and inspiring survey of Identity-Based ring
%signatures from a number of perspectives. 
Some ring signature
schemes with constant-size were also presented in \cite{Y. Dodis,L.
Nguyen}.
%Threshold ring signatures were studied by Bresson et al.
%\cite{Bresson} and Wong et al. \cite{Duncan}.
%Other variations of
%basic ring signatures such as linkable ring signature \cite{J.
%Liu ww}, blind ring signatures \cite{Chan} were also introduced.\\

\indent In terms of security models for provably secure ring
signature schemes, there are three models commonly used. They
provide different security levels. The first and the weakest model
was introduced by Rivest et al. \cite{R. Rivest}. Later Abe et al.
\cite{M. Abe} proposed a very strong model. Finally, Liu and Wong
\cite{J. Liu} presented a model whose security level is considered
to be lying in between the two foregoing models. We mainly use the
ideas of constructing IDRS schemes in \cite{Herranz}, and the
security models of ring signatures in \cite{J. Liu} in this paper.\\
%%%%%%%%%%%%%%%%%%%%%%%%%%%%%%%%%%%%%%%%%%%%%%%%%%%
%                                                 %
%      Certificateless Signatures                 %
%                                                 %
%%%%%%%%%%%%%%%%%%%%%%%%%%%%%%%%%%%%%%%%%%%%%%%%%%%
\indent CL-PKC has got fruitful achievements since its introduction
in \cite{Al-Riyami,CZQWZ09,CZQWZ10,DZL14,YZHMSZ10,agg3,agg4}. Al-Riyami and Paterson presented
\cite{Al-Riyami} the first certificateless signature (CLS) scheme.
Since then, several CLS schemes \cite{Hu,X.
Li,MZZ10a,MZZ10b,Yap,Yum,ZWDQ10,Z. Zhang} were proposed. In \cite{Huang}, Huang et
al. defined the security model of CLS schemes. Zhang et al. \cite{Z.
Zhang} improved the security model of CLS schemes, and presented a
secure CLS scheme. Generic ways to construct CLS schemes were
investigated in \cite{Yum}, \cite{Hu}. In \cite{X. Li}, a
certificateless proxy signature scheme was proposed.
% A generic
%construction of CLE secure in the standard model was given in
%\cite{Sherman}, while \cite{J. Liu AS} gave a specific construction
%of CLE and the first CLS scheme secure in the standard model.
An work about certificateless ring signature was done by
Chow and Yap \cite{S. Chow3}. The security of their scheme is based
on the hardness of the $k$-CAA problem and Modified Inverse
Computational Diffie-Hellman problem and is proved in a weak model
that requires a type I adversary to submit the secret values
corresponding to the replaced public keys to the challenger in the
sign queries. The computional cost of their scheme involves a large
amount of paring operations which linearly increase with the number
of group members. %So far as we know, there is no certificateless
%ring signature (CL-Ring) scheme whose security is based on some
%classical assumptions.

\section{Preliminaries}
In this section, we will review some fundamental backgrounds
required in this paper.

\subsection{Bilinear Pairings and Computational
Problems}\label{Sub-Pairing} Let $G_1$ be an additive group of prime
order $q$ and $G_2$ be a multiplicative group of the same order. Let
\emph{P} denote a generator of $G_1$. A mapping $e:G_1\times
G_1\longrightarrow G_2$ is called a bilinear mapping if it satisfies
the following properties:
\begin{enumerate}
  \item Bilinear: $e(aP,bQ)=e(P,Q)^{ab}$ for all $P,Q\in G_1, a, b\in Z_q^*$.
  \item Non-degeneracy: There exists $P,Q\in G_1$ such that $e(P, Q)\neq 1$.
  \item Computable: There exists an efficient algorithm to compute $e(P,Q)$ for any $P,Q\in
  G_1$.
\end{enumerate}

For a group \emph{G} of prime order, we denote the set
$G^*=G\setminus \left\{\mathcal {O}\right\}$, where $\mathcal {O}$
is the identity element of the group.

\medskip
\noindent {\bf Discrete Logarithm (DL) Problem:} \emph{Given a
generator $P$ of a cyclic additive group $G$ with order $q$, and
$Q\in G^*$ to find an integer $a\in Z_q^*$ such that $Q=aP$.}

\medskip
\noindent{\bf Computational Diffie-Hellman (CDH) Problem:}
\emph{Given a generator $P$ of a cyclic additive group $G$ with
order $q$, and given $(aP,bP)$ for unknown $a,b\in Z_q^*$; to
compute $abP$.}

\subsection{The Concept of Certificateless Ring Signature Schemes}
A CL-Ring scheme is defined by seven algorithms: {\sf Setup}, {\sf
Partial-Private-Key-Extract}, {\sf Set-Secret-Value}, {\sf
Set-Private-Key}, {\sf Set-Public-Key}, {\sf Ring-Sign} and {\sf
Verify}. The description of each algorithm is as follows.
\begin{itemize}
\item  {\sf Setup}: This algorithm runs by the KGC that takes as input a
security parameter $\ell$ to produce a {\sf masterkey} and a list of
system parameters {\sf param}.

\item {\sf Partial-Private-Key-Extract}: This algorithm runs by the KGC that takes as input a
user's identity $ID$, a parameter list {\sf param} and a {\sf
masterkey} to produce the user's partial private key $D_{ID}$.

\item {\sf Set-Secret-Value}: This algorithm takes as input a parameter list
{\sf param} and a user's identity $ID$ to produce the user's secret
value $x$.

\item {\sf Set-Private-Key}: This algorithm takes as input a parameter list
{\sf param}, a user's identity $ID$, the user's partial private key
$D_{ID}$ and secret value $x$ to produce a private signing key
$S_{ID}$ for this user.

\item {\sf Set-Public-Key}: This algorithm takes as input a parameter list
{\sf param}, a user's identity $ID$ and secret value $x_{ID}$ to
produce a public key $P_{ID}$ for the user.

\item {\sf Ring-Sign}: This algorithm takes as input a message $M\in \mathcal {M},\mathcal {M}$ is
the message space, a set of \emph{n} group members whose identities
form the set $L_{ID}=\{ID_1,...,ID_n\}$ and their corresponding
public keys form the set $L_{PK}=\{P_{ID_1},...,P_{ID_n}\}$, a
parameter list {\sf param} and a singer's signing key $S_{ID_s}$ to
produce a ring signature $\sigma$. Here $S_{ID_s}$ is the $s$-th
group member's private key.

\item {\sf Verify}: This algorithm takes as input a message $M$, a ring signature $\sigma$, a parameter
list {\sf param}, the set $L_{ID}$ of the group members' identities
and the set $L_{PK}$ of the corresponding public keys of the group
members to output $True$ if the signature is correct, or $False$
otherwise.
\end{itemize}

\section{Security Models of Certificateless Ring Signature
Schemes}\label{Sub-Adver-Model} There are two types of adversaries
in the certificateless system: namely \emph{Type I Adversary} and
\emph{Type II Adversary}. A \emph{Type I Adversary} $\mathcal {A}_I$
simulates attacks when the adversary (anyone except the KGC)
replaces the user's public key with a value of his/her choice.
However, $\mathcal {A}_I$ is not given this user's partial private
key $D_{ID}$ (and system's \textsf{masterkey}). On the other hand, a
\emph{Type II Adversary} $\mathcal {A}_{II}$ has access to the {\sf
masterkey} but cannot perform public key replacement.

Combining the security notions of certificateless public key
cryptography and traditional ring signature schemes, we define the
security of a CL-Ring scheme via the following two games between a
challenger $\mathcal {C}$ and an adversary $\mathcal {A}_I$ or
$\mathcal {A}_{II}$.

\medskip
\noindent \textbf{Game 1: Unforgeability of CL-Ring against Type I
Adversary $\mathcal{A}_I$}

\medskip
\noindent {\em Setup}: $\mathcal {C}$ runs the {\sf Setup}
algorithm, takes as input a security parameter $\ell$ to obtain a
{\sf masterkey} and the system parameter  {\sf param}. $\mathcal
{C}$ then sends {\sf param} to the adversary $\mathcal {A}_I$ while
keeping the {\sf masterkey} as secret. In addition, $\mathcal{C}$
will maintain three lists $L_1,L_2,L_3$ where
\begin{itemize}
\item $L_1$ is used to record the identities which have been chosen by $\mathcal{A}_I$
in the {\sf Partial-Private-Key Queries}.

\item $L_2$ is used to record the identities whose public keys have
been replaced by $\mathcal{A}_I$.

\item $L_3$ is used to record the identities which have been chosen by $\mathcal{A}_I$
in the  {\sf Private-Key Queries}.

\end{itemize}
All these three lists $L_1,L_2,L_3$ are the empty set $\emptyset$ at
the beginning of the game.

\medskip
\noindent \emph{Training}: The adversary $\mathcal {A}_I$ can
adaptively  issue a polynomially bounded number of queries as
defined below:
\begin{itemize}
\item {\sf Partial-Private-Key Queries} $PPK(ID)$: $\mathcal {A}_I$ can request the
partial private key of any user whose identity is $ID$. In respond,
\begin{enumerate}
\item $\mathcal{C}$ first resets $L_1=L_1\cup\{ID\}$.

\item $\mathcal {C}$ then runs the algorithm {\sf
Partial-Private-Key-Extract} and outputs the partial private key
$D_{ID}$.
\end{enumerate}

\item {\sf Public-Key Queries} $PK(ID)$: $\mathcal{A}_I$ can request the
public key of a user whose identity is $ID$. In respond,

\begin{enumerate}
\item $\mathcal {C}$ first runs the algorithm {\sf Set-Secret-Value}
and obtains the secret value $x_{ID}$.

\item $\mathcal{C}$ then runs
the algorithm {\sf Set-Public-Key} and obtains the public key
$P_{ID}$. $\mathcal{C}$ outputs the public key $P_{ID}$ as the
answer.
\end{enumerate}

\item {\sf Public-Key-Replacement Queries} $PKR(ID, P_{ID}')$: For any user whose identity is
$ID$, $\mathcal {A}_I$ can choose a new public key $P_{ID}'$.
$\mathcal {A}_I$ then sets $P_{ID}'$ as the new public key of this
user and submits ($ID,P_{ID}'$) to $\mathcal {C}$. On receiving a
query $PKR(ID, P_{ID}')$, $\mathcal{C}$ resets $L_2=L_2\cup\{ID\}$
and updates the public key of this user to the new value $P_{ID}'$.

\item {\sf Private-Key Queries} $PrK(ID)$: $\mathcal{A}_I$ can request the
private key of a user whose identity is $ID$. In respond,
\begin{enumerate}
\item $\mathcal {C}$ first checks the set $L_2$. If $ID\in L_2$ (that is, the public key of the user $ID$ has
been replaced), $\mathcal{C}$ will return the symbol $\bot$ which
means $\mathcal{C}$ cannot output the private key of an identity
whose public key has been replaced.

\item  Otherwise, $ID \notin L_2$ and $\mathcal{C}$ resets $L_3=L_3\cup\{ID\}$.
$\mathcal{C}$ then runs the algorithm {\sf Set-Private-Key} and
outputs the private key $S_{ID}$.
\end{enumerate}

\item {\sf Ring-Sign Queries} $RS(M, L_{ID},L_{PK})$: $\mathcal{A}_I$ can request the ring signature
of a message $M$ on behalf of a group whose identities are listed in
the set $L_{ID}$ and the corresponding public keys are in the set
$L_{PK}$. In respond, $\mathcal {C}$ outputs a ring signature
$\sigma$ for the message $M$. It is required that the algorithm {\sf
Verify} will output $True$ for the input $(M,\sigma,{\sf param},
L_{ID}, L_{PK})$.
\end{itemize}

\noindent \emph{Forgery}: Finally, $\mathcal{A}_I$ outputs a tuple
$(M^*,\sigma^*,L_{ID}^*,L_{PK}^*)$ as the forgery. We say
$\mathcal{A}_{I}$ wins the game if the forgery satisfies all the
following requirements:
\begin{enumerate}
\item  The algorithm {\sf Verify} outputs $True$ for the input $(M^*,\sigma^*,{\sf param},
L_{ID}^*, L_{PK}^*)$.

\item $L_{ID}^* \cap L_1 \cap L_2 =\emptyset$ and $L_{ID}^*\cap L_3=\emptyset$.

\item $(M^*,L_{ID}^*,L_{PK}^*)$ has never been queried during the {\sf Ring-Sign Queries}.
\end{enumerate}

\medskip
\noindent\textbf{Game 2: Unforgeability of CL-Ring against Type II
Adversary $\mathcal{A}_{II}$}

\medskip
\noindent {\em Setup}: $\mathcal {C}$ runs the {\sf Setup}
algorithm, takes as input a security parameter $\ell$ to obtain the
system parameter list {\sf param} and also the system's {\sf
masterkey}. $\mathcal {C}$ then sends {\sf param} and {\sf
masterkey} to the adversary $\mathcal {A}_{II}$. $\mathcal{C}$ will
maintain two lists $L_1,L_2$ where
\begin{itemize}
\item $L_1$ is used to record the identities whose public keys have
been replaced by $\mathcal{A}_{II}$.

\item $L_2$ is used to record the identities which have been chosen by $\mathcal{A}_{II}$
in the  {\sf Private-Key Queries}.

\end{itemize}
Both two lists $L_1,L_2$ are  empty  at the beginning of the game.

\medskip
\noindent \emph{Training}: As defined in \textbf{Game 1}, the type
II adversary $\mathcal {A}_{II}$ can issue a polynomially bounded
number of {\sf Public Key Queries}, {\sf Private-Key Queries}, {\sf
Public-Key-Replacement Queries} and {\sf Ring-Sign Queries}.
$\mathcal{C}$ will answer those queries as same in \textbf{Game 1}.
Note that $\mathcal{A}_{II}$ does not need to issue
Partial-Private-Key queries because he has already known the
system's {\sf masterkey}.

\medskip \noindent \emph{Forgery}: Finally, $\mathcal{A}_{II}$
outputs a tuple $(M^*,\sigma^*,L_{ID}^*,L_{PK}^*)$ as the forgery.
We say $\mathcal{A}_{II}$ wins the game if the forgery satisfies all
the following requirements:
\begin{enumerate}

\item  The algorithm {\sf Verify} outputs $True$ for the input $(M^*,\sigma^*,{\sf param},
L_{ID}^*, L_{PK}^*)$.

\item $L_{ID}^* \cap L_1 =\emptyset$ and $L_{ID}^*\cap L_2=\emptyset$.

\item $(M^*,L_{ID}^*,L_{PK}^*)$ has never been queried during the {\sf Ring-Sign Queries}.

\end{enumerate}

\begin{define}
A CL-Ring scheme is existentially unforgeable under adaptively
chosen-message attack \emph{iff} the success probability of any
polynomially bounded adversary in the above two games is negligible.
\end{define}

\begin{define}
A CL-Ring scheme is said to have the unconditional signer anonymity
if for any group of $n$ users whose identities form the set $L_{ID}$
and their corresponding public keys form the set $L_{PK}$, any
message $M$ and any ring signature $\sigma=${\sf
$Ring$-$Sign$}$(M,L_{ID},L_{PK},S_{ID_s})$, any verifier $\mathcal
{V}$ cannot identify the actual signer with probability better than
a random guess. That is, $\mathcal {V}$ can only output the actual
signer with probability no better than $\frac{1}{n}$
($\frac{1}{n-1}$ when $\mathcal {V}$ is in the signers' ring).
\end{define}

\section{Analysis of A Generic Construction of
CL-Ring}\label{Sec-Analysis-Generic-Construction} In \cite{Yum}, Yum
and Lee presented a generic way to construct a certificateless
signature scheme. However, Hu {\em et al.} \cite{Hu} pointed out
that their construction is flawed and proposed a new one. It seems
at first glance that the methods in \cite{Hu} can also be used to
obtain a generic construction of CL-Ring signatures. However, as we
will show later, the resulting scheme is not secure in our security
model defined in Section~\ref{Sub-Adver-Model}.

\subsection{A Generic Construction of CL-Ring}
Let $\prod_{PK}=(Gen_{PK},Ring$-$Sign_{PK},Ver_{PK})$ be a
traditional public key-based ring signature scheme which is
existentially unforgeable under adaptively chosen-message attack.
$Gen_{PK}$ takes a security parameter as input and generates a
public/secret pair $(pk_{PK},sk_{PK})$; $Ring$-$Sign_{PK}$ takes a
private signing key, a set of public keys and a message as inputs,
and generates a ring signature $\sigma_{PK}$; and $Ver_{PK}$ is the
corresponding ring signature verification algorithm.

Let $\prod_{ID}=(Gen_{ID},KGen_{ID},Ring$-$Sign_{ID},Ver_{ID})$ be
an identity-based ring signature scheme that is existentially
unforgeable under adaptively chosen-message and identities attacks.
$Gen_{ID}$ takes a security parameter as input and generates a
master secret key {\sf masterkey} and a list of system parameters
{\sf param}; $KGen_{ID}$ is an identity-based secret key generation
algorithm which takes {\sf masterkey} and an identity $ID$ and
generates a secret key denoted by $D_{ID}$; $Ring$-$Sign_{ID}$ takes
a private signing key, a set of identities and a message as inputs,
and generates an identity-based ring signature denoted by
$\sigma_{ID}$; and $Ver_{ID}$ is the corresponding ring signature
verification algorithm.

As defined in Section~\ref{Sub-Adver-Model}, a CL-Ring signature
scheme consists of seven algorithms. Using the similar methods in
\cite{Hu},  we can obtain a generic construction of CL-Ring as
described in Fig~\ref{Fig-Generic-Construction}.
\begin{figure}\caption{A Generic Construction of CL-Ring}
\begin{boxit}
\begin{description}
\item {\sf Setup}: On input $1^\ell$, the KGC runs $Gen_{ID}(1^\ell)$ to produce a {\sf masterkey} and a list of
system parameters {\sf param}.
\item {\sf Partial-Private-Key-Extract}: This algorithm accepts {\sf param}, {\sf masterkey}, a user's identity
$ID\in \{0, 1\}^*$ and runs $KGen_{ID}(param, masterkey,ID)$ to
output the user's partial private key $D_{ID}$.
\item {\sf Set-Secret-Value}: This algorithm accepts {\sf param}, a user's identity
$ID$ and runs $Gen_{PK}(1^\ell)$ to generate a public/secret
$(P_{ID},x_{ID})$ pair and returns $x_{ID}$ as user's secret value .
\item {\sf Set-Private-Key}: This algorithm takes as input a parameter list
{\sf param}, a user's identity $ID$, the user's partial private key
$D_{ID}$ and secret value $x_{ID}$ to produce a private signing key
$S_{ID}=(x_{ID},D_{ID})$ for this user.
\item {\sf Set-Public-Key}: This algorithm takes as input a parameter list
{\sf param}, a user's identity $ID$, $(P_{ID},x_{ID})$ and outputs
public key $P_{ID}$ for the user.
\item {\sf Ring-Sign}: This algorithm takes as input a message $M\in \mathcal {M}$,
  a set of \emph{n} group members whose identities
form the set $L_{ID}=\{ID_1,...,ID_n\}$ and their corresponding
public keys form the set $L_{PK}=\{P_{ID_1},...,P_{ID_n}\}$, a
parameter list {\sf param} and a singer's signing key
$S_{ID_s}=(x_{ID_s},D_{ID_s})$. Here $x_{ID_s}$ and $D_{ID_s}$ is
the $s$-th group member's secret value and partial private key
respectively. To generate a ring signature $\sigma$, the signer does
the following.
\begin{itemize}
  \item Set $M'=M||param||L_{ID}||L_{PK}$;
  \item Compute $\sigma_{PK}=Ring$-$Sign_{PK}(x_{ID_s},L_{PK},M')$;
  \item Set $M''=M||param||L_{ID}||L_{PK}||\sigma_{PK}$;
  \item Compute $\sigma_{ID}=Ring$-$Sign_{ID}(D_{ID_s},L_{ID},M'')$;
  \item Set $\sigma=\sigma_{PK}||\sigma_{ID}$.
\end{itemize}
\item {\sf Verify}: If $Ver_{PK}(\sigma_{PK},L_{PK},M')=True$ and
  $Ver_{ID}(\sigma_{ID},L_{ID},M'')=True$ then the algorithm outputs
  $True$, otherwise outputs $False$.
\end{description}
\end{boxit}\label{Fig-Generic-Construction}
\end{figure}

\subsection{Security Analysis of the Generic Construction}
In this section, we will show that the generic construction
described in Fig~\ref{Fig-Generic-Construction} is not secure under
the definition in Section~\ref{Sub-Adver-Model}.

We firstly show that a type I adversary $\mathcal{A}_I$ can forge a
valid ring signature of any message $M$. The attack algorithm is
described as below:

\begin{itemize}
\item $\mathcal{A}_I$ first chooses $n$ identities
$(ID_1,ID_2,\cdots,ID_n)$ and sets
$$L_{ID}=\{ID_1,ID_2,\cdots,ID_n\}.$$

\item As defined in the Game 1 in Section~\ref{Sub-Adver-Model}, $\mathcal{A}_I$ then issues $n$ Public-Key queries to obtain the
corresponding public keys $(P_{ID_1},P_{ID_2},\cdots,P_{ID_n})$.

\item $\mathcal{A}_I$ runs the algorithm {\sf
Set-Secret-Value} to generate a secret value $x_{ID_i}$ for the user
$ID_i \in L_{ID}$. It also runs the algorithm {\sf Set-Public-Key}
to obtain a public key $P_{ID_i}'$. Finally, it replaces $ID_i$'s
public key with $P_{ID_i}'$ and sets
$$L_{PK}=\{P_{ID_1},P_{ID_2},\cdots, P_{ID_i}', \cdots, P_{ID_n}\}.$$

\item $\mathcal{A}_I$ then submits a partial private key query
for an identity $ID_j \in L_{ID}$ and obtains the partial private
key $D_{ID_j}$, with the only requirement that $ID_j\neq ID_i$.

\item For any message $M$, $\mathcal{A}_I$ sets $M'=M||param||L_{ID}||L_{PK}$ and uses $x_{ID_i}$ to compute
 $$\sigma_{PK}=Ring\mathrm{-}Sign_{PK}(x_{ID_i},L_{PK},M').$$
\item It then sets $M''=M||param||L_{ID}||L_{PK}||\sigma_{PK}$ and uses $D_{ID_j}$ to compute
$$\sigma_{ID}=Ring\mathrm{-}Sign_{ID}(D_{ID_j},L_{ID},M'')$$

\item $\mathcal{A}_{I}$ outputs $(M,
\sigma=\sigma_{PK}||\sigma_{ID},L_{ID},L_{PK})$ as the forgery.
\end{itemize}
As we can see, $\sigma$ is a valid ring signature of $M$ under
$L_{ID}$ and $L_{PK}$. This is because $\mathcal{A}_{I}$ runs all
the algorithms as same as defined in the generic construction in
Section~\ref{Sub-Adver-Model}. We note that this attack is a strong
attack that belongs to the no-message attack classes, where no
signing oracle is required.

The generic construction given in Section~\ref{Sub-Adver-Model} only
guarantees that the singer of a valid ring signature possesses a
secret value $x_{ID_i}$ of a user $ID_i \in L_{ID}$ and a partial
private key $D_{ID_j}$ of a user $ID_j \in L_{ID}$, instead of
proving that the signer must know the private key of one user (i.e.,
$ID_i=ID_j$). This is the reason why a Type I adversary can forge a
valid signature for any message. How to give a provably secure
generic construction of certificateless ring signature is still an
open problem.

\section{A Concrete Certificateless Ring Signatures
Scheme}\label{Sec-Scheme} In this section, we will give the concrete
construction of certificateless ring signature.

\subsection{Description of Our CL-Ring Scheme}
Our CL-Ring scheme consists of the following concrete algorithms:

\begin{itemize}
  \item {\sf Setup}: Given a security parameter $\ell$, the algorithm works as follows.
  \begin{enumerate}
    \item Specify $G_1,G_2, e$, as described in
    Section~\ref{Sub-Pairing}.
    \item Arbitrarily choose a generator $P\in G_1$ and set $g=e(P,P)$.
    \item Choose a random {\sf masterkey} $\kappa\in Z_q^*$ and set $P_0=\kappa P$.
    \item Choose cryptographic hash functions $H_1:\{0,
1\}^*\longrightarrow G_1$, $H_2:\{0, 1\}^*\longrightarrow Z_q^* $
and $H_3:\{0,1\}^*\longrightarrow G_1$.
  \end{enumerate}
  The system parameters
{\sf param}=($G_1,G_2,e, P,g,P_0,H_1,H_2,H_3$). The message
    space is $\mathcal {M}$$= \left\{0,1\right\}^*$.
  \item {\sf Partial-Private-Key-Extract}: This algorithm accepts {\sf param}, {\sf masterkey} and a user's identity
    $ID_i\in \{0, 1\}^*$ to output the user's partial private key $D_i =\kappa Q_i$.
    Where $Q_i=H_1(ID_i)$.
  \item {\sf Set-Secret-Value}:  Given {\sf param}, this algorithm selects a random $x_i\in Z_q^*$ as
    the user's (whose identity is $ID_i$) secret value.
  \item {\sf Set-Private-Key}: This algorithm takes as input {\sf param},
  a user's identity $ID_i$, the user's partial private key $D_i$ and the user's secret value $x_i\in
    Z_q^*$. The output of the algorithm is the user's private key $S_i=(x_i,D_i)$.
  \item {\sf Set-Public-Key}: This algorithm accepts
  {\sf param},
    a user's identity $ID_i$ and his secret value $x_i\in Z_q^*$ to produce the user's public key
    $P_i=x_iP$.
  \item {\sf Ring-Sign}: Suppose there's a group of \emph{n} users whose identities
  form the set
  $L_{ID}=\{ID_1,...,ID_n\}$, and their corresponding public keys form
  the set
  $L_{PK}=\{P_1,...,P_n\}$. To sign a message $M\in \mathcal
  {M}$ on behalf of the group, the actual signer, indexed by \emph{s} using the private key $S_s=(x_s,D_s)$,
    performs the following steps.
    \begin{enumerate}
      \item For each $i\in\{1,...,n\}\backslash\{s\}$,  select $r_i\in
      Z_q^*$ uniformly at random, compute $y_i=g^{r_i}$.
      \item Compute $h_i=H_2(M||L_{ID}||L_{PK}||y_i)$ for all
      $i\in\{1,...,n\}\backslash\{s\}$.
      \item Choose random $r_s\in Z_q^*$, compute $U=H_3(M||L_{ID}||L_{PK})$, $y_s=g^{r_s}e(-P_0,\sum_{i\neq s}h_iQ_i)e(-U,$ $\sum_{i\neq s}h_iP_i)$.
       If $y_s=1_{G_2}$ or $y_s = y_i$ for some $i\neq s$, then redo this step.
      \item Compute $h_s=H_2(M||L_{ID}||L_{PK}||y_s)$.
      \item Compute $V=(\sum_{i=1}^nr_i)P+h_s(D_s+x_sU)$.
      \item Output the ring signature on $M$ as $\sigma=\{(y_1,...,y_n),V\}$.
    \end{enumerate}
  \item {\sf Verify}: To verify a ring signature $\sigma=\{(y_1,...,y_n),V\}$ on a message
  $M$ with identities in $L_{ID}$ and corresponding public keys in $L_{PK}$, the verifier performs the following steps.
    \begin{enumerate}
      \item Compute $h_i=H_2(M||L_{ID}||L_{PK}||y_i)$ for all
      $i\in\{1,...,n\}$, compute $U=H_3(M||L_{ID}||L_{PK})$.
      \item Verify
      $e(V,P)\stackrel{?}{=}y_1\cdot...\cdot y_ne(\sum_{i=1}^nh_iQ_i,P_0)e(\sum_{i=1}^nh_iP_i,U)$
     holds with equality.
      \item Accept the ring signature as valid and output $True$ if the above equation holds, otherwise, output $False$.
    \end{enumerate}
\end{itemize}

\subsection{Efficiency}
We only consider the costly operations including the pairing
operation (Pairing), scalar multiplication in $G_1$ ($G_1$ SM),
exponentiation in $G_2$ ($G_2$ E) and MapToPoint hash operation
\cite{agg4} (Hash). The numbers of these operations in our scheme are
shown in \textbf{Table 1}.
\begin{center}
\textbf{Table 1.} Efficiency\\
\begin{tabular}
{|c|c|c|c|c|}
  \hline
   &Pairing &$G_1$ SM &$G_2$ E& Hash  \\
  \hline
  Sign  &2 & 2n+3 &n &n+1  \\
  \hline
  Verify &3 & 2n &0 &n+1 \\
  \hline
  Total&5 &4n+3 &n & 2n+2\\
  \hline
\end{tabular}
\end{center}
Pairing operation is the most time consuming operation. Our CL-Ring
scheme only requires 5 pairing operations which is independent of
the group size.

\section{Analysis of the Proposed CL-Ring Scheme}\label{Sec-Proof}
In this section, we will analyze our proposed scheme in detail.
\subsection{Correctness}
The correctness of the proposed scheme can be easily verified with
the following:
\begin{eqnarray*}
&&e(V,P)=e((\sum_{i=1}^nr_i)P+h_s(D_s+x_sU),P)\\
&=&e((\sum_{i=1}^nr_i)P,P)e(h_s(D_s+x_sU),P)\\
&=&y_1\cdot...\cdot y_ne(\sum_{i\neq s}h_iQ_i,P_0)e(\sum_{i\neq s}h_iP_i,U)e(h_sD_s,P)e(h_sx_sU,P)\\
&=&y_1\cdot...\cdot y_ne(\sum_{i\neq s}h_iQ_i,P_0)e(\sum_{i\neq s}h_iP_i,U)e(h_sQ_s,P_0)e(h_sP_s,U)\\
&=&y_1\cdot...\cdot y_ne(\sum_{i=1}^nh_iQ_i,P_0)e(\sum_{i=1}^nh_iP_i,U)\\
\end{eqnarray*}
\subsection{Unconditional Anonymity}
Let $\sigma=\{(y_1,...,y_n),V\}$ be a valid ring signature of a
message $M$ on behalf of a group of $n$ members specified by
identities in $L_{ID}$ and public keys in $L_{PK}$. Since all the
$r_i,i\in\{0,...,n\}\backslash \{s\}$ are randomly generated, hence
all $y_i,i\in\{0,...,n\}\backslash \{s\}$ are also uniformly
distributed. The randomness of $r_s$ chosen by the signer implies
$y_s=g^{r_s}e(-P_0,\sum_{i\neq s}h_iQ_i)e(-U,\sum_{i\neq s}h_iP_i)$
is also uniformly distributed. So $(y_1,...,y_n)$ in the signature reveals no information about the signer.\\
\indent It remains to consider whether
$V=(\sum_{i=1}^nr_i)P+h_s(D_s+x_sU)$ leaks information about the
actual signer. From the construction of $V$, it is obvious to see
that $D_s+x_sU=h_s^{-1}(V-(\sum_{i=1}^nr_i)P)$. To identify whether
$ID_s$ is the identity of the actual signer, the only way is to
check $e(Q_s,P_0)e(P_s,U)\stackrel{?}{=}e(D_s+x_sU,P)$. Namely,
$e(Q_s,P_0)e(P_s,U)\stackrel{?}{=}e(h_s^{-1}(V-(\sum_{i=1}^nr_i)P),P)$.
If $ID_s$ is the identity of the actual signer, it should hold
$$y_s=g^{r_s}e(-P_0,\sum_{i\neq s}h_iQ_i)e(-U,\sum_{i\neq s}h_iP_i).$$
It remains to check
$$e(Q_s,P_0)e(P_s,U)\stackrel{?}{=}(\frac{e(V,P)}{y_1\cdot...\cdot y_ne(P_0,\sum_{i\neq
s}h_sQ_s)e(U,\sum_{i\neq j}h_sP_i)})^{h_s^{-1}}$$ However,we have
for each $j\in \{1,2,...,n\}$
\begin{eqnarray*}
&&(\frac{e(V,P)}{y_1\cdot...\cdot y_ne(P_0,\sum_{i\neq j}h_iQ_i)e(U,\sum_{i\neq j}h_iP_i)})^{h_j^{-1}}\\
&=&(\frac{e(\sum_{i=1}^nr_i)P+h_s(D_s+x_sU),P)}{y_1\cdot...\cdot y_ne(P_0,\sum_{i\neq s}h_iQ_i)e(U,\sum_{i\neq s}h_iP_i)e(h_sQ_s,P_0)e(h_sP_s,U)w})^{h_j^{-1}}\\
&=&(\frac{e((\sum_{i=1}^nr_i)P,P)}{e((\sum_{i\neq s}r_i)P,P)y_se(P_0,\sum_{i\neq s}h_iQ_i)e(U,\sum_{i\neq s}h_iP_i)w})^{h_j^{-1}}\\
&=&w^{-h_j^{-1}}=e(Q_j,P_0)e(P_j,U)
\end{eqnarray*}
where $w=e(-h_jQ_j,P_0)e(-h_jP_j,U)$, and $ID_s$ is the identity of
the actual signer. This fact shows that $V$ in the signature does
not leak any information about the identity of the actual signer.
And hence, the unconditional anonymity of our CL-Ring scheme is
proved.
\subsection{Unforgeability}
Assuming that the CDH problem is hard, we now show the
unforgeability of our CL-Ring scheme.
\begin{theorem}\label{Theorem-1}
In the random oracle model \cite{Bellare}, if $\mathcal {A}_I$ can
win the Game 1, with an advantage $\epsilon\geq
7P_n^{q_{H_1}}/2^\ell$ within a time span $t$ for a security
parameter $\ell$; and asking at most $q_K$ Partial-Private-Key
queries, at most $q_P$ Public-Key queries, at most $q_{Pr}$
Private-Key queries, at most $q_{H_1}$ $H_1$ queries, at most
$q_{H_2}$ $H_2$ queries, at most $q_{H_3}$ $H_3$ queries, $q_S$
Ring-Sign queries. Then the CDH problem in $G_1$ can be solved
within time
$2(t+q_{H_1}T_1+q_{H_2}T_2+q_{H_3}T_3+q_KT_K+q_PT_P+q_{Pr}T_{Pr}+q_ST_S)$
and with probability $\geq
((\frac{q_K+q_{Pr}}{q_K+q_{Pr}+n})^{q_K+q_{Pr}+n}(\frac{n}{q_K+q_{Pr}+n})^n\epsilon)^2/66P_n^{q_{H_1}}$
where n is the ring scale, $P_n^{q_{H_1}}$ is defined as the number
of $n$-permutations of $q_{H_1}$ elements i.e.
$P_n^{q_{H_1}}=q_{H_1}\cdot...\cdot(q_{H_1}-n+2)\cdot(q_n-n+1)$,
$T_1$ (resp. $T_2,T_3,T_K,T_P,T_{Pr}$ and $T_S$) is the time cost of
an $H_1$ (resp. $H_2,H_3$, Partial-Private-Key, Public-Key,
Private-Key and Ring-Sign) query.
\end{theorem}

Please refer to Appendix~\ref{Appendix-Proof-Th1}.

\begin{theorem}\label{Theorem-2}
In the random oracle model, if $\mathcal {A}_{II}$ can win the Game
2, with an advantage $\epsilon\geq 7P_n^{q_{H_1}}/2^\ell$ within a
time span $t$ for a security parameter $\ell$; and asking at most
$q_P$ Public-Key queries, at most $q_{K}$ Private-Key queries, at
most $q_{H_1}$ $H_1$ queries, at most $q_{H_2}$ $H_2$ queries, at
most $q_{H_3}$ $H_3$ queries, at most $q_S$ Ring-Sign queries. Then
the CDH problem in $G_1$ can be solved within time
$2(t+q_{H_1}T_1+q_{H_2}T_2+q_{H_3}T_3+q_KT_{Pr}+q_PT_P+q_ST_S)$ and
with probability $\geq
((\frac{q_K}{q_K+n})^{q_K+n}(\frac{n}{q_K+n})^n
\epsilon)^2/66P_n^{q_{H_1}}$.
\end{theorem}

Please refer to Appendix~\ref{Appendix-Proof-Th2}.

\section{Conclusion}\label{Sec-Conclusion}
In this paper, we proposed a concrete construction of
certificateless ring signature scheme from the bilinear pairing. The
security models of certificateless ring signatures are also
formalized. The models capture the essence of the possible
adversaries in the notion of certificateless system and ring
signatures. In the random oracle models, the unforgeability of our
scheme is based on the hardness of Computational Diffie-Hellman
problem. We note that the number of pairing computation in our
scheme is constant and does not grow with the number of group
members.

\appendix
\section{Proof of Theorem~\ref{Theorem-1}}\label{Appendix-Proof-Th1}
\emph{Proof}. Let $\mathcal {C}$ be a CDH attacker, $\mathcal {A}_I$
be a type I adversary of our CL-Ring scheme who interacts with
$\mathcal {C}$ following Game 1 and can forge a valid ring
signature. Suppose $\mathcal {C}$ receives a random instance
$(P,aP,bP)$ of the CDH problem in $G_1$. We show how $\mathcal {C}$
can use $\mathcal {A}_I$ to solve the CDH problem, i.e. to compute
$abP$.

\medskip
\noindent \emph{Setup:} $\mathcal {C}$ first sets $P_0=aP$ and
selects {\sf param}=$(G_1,G_2,e,P,g,P_0,H_1,H_2,H_3)$, then sends
{\sf param} to $\mathcal {A}_I$. We take hash functions $H_1,H_2$
and $H_3$ as random oracles.

\medskip
\noindent \emph{Training:} $\mathcal {A}_I$ can ask $\mathcal {C}$
$H_1,H_2,H_3$, Partial-Private-Key, Public-Key, Private-Key,
Public-Key-Replacement and Ring-Sign queries. In order to maintain
consistency and avoid conflict, $\mathcal {C}$ keeps four lists
${\bf H_1}$, ${\bf H_2}$, ${\bf H_3}$, and \textbf{K} to store the
answers used, where ${\bf H_1}$ includes items of the form
$(ID,\alpha,Q_{ID},c)$, ${\bf H_2}$ includes items of the form
$(M,L_{ID},L_{PK},y,h)$, ${\bf H_3}$ includes items of the form
$(M,L_{ID},L_{PK},\beta,U,c'')$, and \textbf{K} includes items of
the form $(ID,x,D_{ID},P_{ID},c')$. All of these four lists are
initially empty. $\mathcal{C}$ also maintains three lists
$L_1,L_2,L_3$, the function of these three lists are the same as
mentioned in Game 1 Section~\ref{Sub-Adver-Model}.

\medskip
\noindent $H_1$ \emph{Queries}: On receiving a query $H_1(ID)$,
$\mathcal {C}$ does as follows.
\begin{enumerate}
  \item If there exists an item $(ID,\alpha,Q_{ID},c)$ in $\bf H_1$, then $\mathcal {C}$ returns $Q_{ID}$ as answer.
  \item Otherwise, $\mathcal {C}$ first flips a coin $c\in \{0,1\}$
    that yields 0 with probability $\delta$ and 1 with
    probability $1-\delta$ ($\delta$ will be determined later), then picks a random element $\alpha$ (has not been used before) in
     $Z_q^*$. If
    $c=0$, $\mathcal {C}$ computes $Q_{ID}=H_1(ID)=\alpha P$; otherwise $c=1$, it computes $Q_{ID}=H_1(ID)=\alpha bP$.
     $\mathcal {C}$  then adds $(ID,\alpha,Q_{ID},c)$ to $\bf H_1$ and returns $Q_{ID}$ as answer.
\end{enumerate}
\noindent $H_2$ \emph{Queries}: On receiving a query
$H_2(M||L_{ID}||L_{PK}||y)$, $\mathcal {C}$ first checks if there
exists an item $(M,L_{ID},L_{PK},y,h)$ in $\bf H_2$, if so, returns
$h$ as answer. Otherwise, $\mathcal {C}$ picks a random $h\in Z_q^*$
which has not been used in the answers of the former $H_2$
\emph{Queries}, then returns $h$ as answer and adds
$(M,L_{ID},L_{PK},y,h)$ to $\bf H_2$.

\medskip
\noindent $H_3$ \emph{Queries}: On receiving a query
$H_3(M||L_{ID}||L_{PK})$, $\mathcal {C}$ first checks if there
exists an item $(M,L_{ID},L_{PK},\beta,U,c'')$ in $\bf H_3$, if so,
returns $U$ as answer. Otherwise, $\mathcal {C}$ first flips a coin
$c''\in \{0,1\}$ that yields 0 with probability $\delta$ and 1 with
probability $1-\delta$ then picks a random $\beta\in Z_q^*$ which
has not been used in the answers of the former $H_3$ \emph{Queries}.
If $c''=0$, compute $U=\beta P$; while $c''=1$, compute $U=\beta
bP$. In both cases, $\mathcal {C}$ will add
$(M,L_{ID},L_{PK},\beta,U,c'')$ to $\bf H_3$ and return $U$ as
answer.

\medskip
\noindent \emph{Partial-Private-Key Queries}: Whenever $\mathcal
{C}$ receives a query $PPK (ID)$
\begin{enumerate}
  \item If there exists an item $(ID,x,D_{ID},P_{ID},c')$ in
  \textbf{K}, $\mathcal {C}$ does the following:
  \begin{enumerate}
    \item If $D_{ID}\neq \bot$, $\mathcal {C}$ returns $D_{ID}$ as
  answer.
    \item Else, if there's an item $(ID,\alpha,Q_{ID},c)$ exists in $\bf
    H_1$, $\mathcal {C}$ sets $L_1=L_1\cup\{ID\}$, $D_{ID}=\alpha P_0$ and returns $D_{ID}$
    as
    answer when $c=0$; while $c=1$, $\mathcal {C}$ aborts.
    \item Otherwise, $\mathcal {C}$
    first makes an $H_1(ID)$ query to obtain an item $(ID,\alpha,Q_{ID},c)$. If $c=1$, $\mathcal
    {C}$ aborts; while $c=0$, $\mathcal {C}$
     sets $L_1=L_1\cup\{ID\}$, $D_{ID}=\alpha P_0$ and returns $D_{ID}$ as
    answer.
  \end{enumerate}
  \item Otherwise $\mathcal {C}$ does the following:
  \begin{enumerate}
    \item If there exists an item $(ID,\alpha,Q_{ID},c)$ in $\bf H_1$, $\mathcal
    {C}$ sets
$L_1=L_1\cup\{ID\}$,
    computes $D_{ID}=\alpha P_0$, sets $x=\bot,P_{ID}=\bot$, adds $(ID,x,D_{ID},P_{ID},c')$ to \textbf{K}
    and returns $D_{ID}$ as answer when
    $c=0$; while $c=1$ $\mathcal {C}$ aborts.
    \item Otherwise, $\mathcal {C}$ first makes an $H_1(ID)$ query to obtain
    an item $(ID,\alpha,Q_{ID},c)$ in  $\bf H_1$, then proceeds as in (a).
  \end{enumerate}
\end{enumerate}

\medskip
\noindent \emph{Public-Key Queries}: Whenever $\mathcal {C}$
receives a query $PK (ID)$
\begin{enumerate}
  \item If there exists an item $(ID,x,D_{ID},P_{ID},c')$ in
  \textbf{K}, $\mathcal {C}$ does the following:
    \begin{enumerate}
      \item If $P_{ID}\neq \bot$, $\mathcal {C}$ returns $P_{ID}$ as
  answer;
      \item Otherwise, $\mathcal {C}$ first flips a coin $c'\in \{0,1\}$
    that yields 0 with probability $\delta$ and 1 with
    probability $1-\delta$, then picks a random $x\in Z_q^*$. If
    $c'=0$, $\mathcal {C}$ sets $P_{ID}=x P$; otherwise $c=1$, it computes $P_{ID}=x aP$.
     $\mathcal {C}$ then
      updates $(ID,x,D_{ID},P_{ID},c')$ with new values and returns
  $P_{ID}$ as answer.
    \end{enumerate}
  \item Otherwise, $\mathcal {C}$ first flips a coin $c'\in \{0,1\}$
    that yields 0 with probability $\delta$ and 1 with
    probability $1-\delta$, then picks a random $x\in Z_q^*$. If
    $c'=0$, $\mathcal {C}$ sets $P_{ID}=x P$; otherwise $c=1$, it computes $P_{ID}=x
    aP$. $\mathcal {C}$ then sets $D_{ID}=\bot$,
  returns
  $P_{ID}$ as answer and adds $(ID,x,D_{ID},P_{ID},c')$ to \textbf{K}.
\end{enumerate}

\noindent \emph{Public-Key-Replacement Queries}: On receiving a
query $PKR(ID,P'_{ID})$ ($\mathcal {C}$ sets $L_2=L_2\cup\{ID\}$),
$\mathcal {C}$ first makes a $PPK (ID)$ query to obtain an item
$(ID,x,D_{ID},P_{ID},c')$, then sets $x=\bot$,
 $P_{ID}=P'_{ID}$, and updates the item $(ID,x,D_{ID},P_{ID},c')$ in \textbf{K} to record this replacement.

\medskip
\noindent \emph{Private-Key Queries}: Whenever receives a query $PrK
(ID)$, if $ID\in L_2$ $\mathcal {C}$ returns $\bot$, otherwise
\begin{enumerate}
  \item When there exists an item $(ID,x,D_{ID},P_{ID},c')$ in \textbf{K}
    \begin{enumerate}
      \item If $x=\bot$, $\mathcal {C}$ first makes a $PK(ID)$ query. If $c'\neq 1$, $\mathcal {C}$ sets
$L_3=L_3\cup\{ID\}$, returns $(x,D_{ID})$ as
      answer; otherwise $\mathcal {C}$ aborts.
      \item Else if $D_{ID}=\bot$, $\mathcal {C}$ first makes a
      $PPK(ID)$ query, if $\mathcal {C}$ does not abort and $c'\neq 1$
      then sets
$L_3=L_3\cup\{ID\}$ and $(x,D_{ID})$ will be returned as answer.
Otherwise $\mathcal {C}$ aborts.
      \item Otherwise, when $c'=1$ $\mathcal {C}$ aborts, while $c'=0$ $\mathcal {C}$ sets
      $L_3=L_3\cup\{ID\}$ and
      returns $(x,D_{ID})$ as answer.
    \end{enumerate}
  \item Otherwise, $\mathcal {C}$ first makes $PK(ID)$ and $PPK(ID)$
  queries. If $\mathcal {C}$ does not abort and $c'\neq 1$,
  then sets
$L_3=L_3\cup\{ID\}$, returns $(x,D_{ID})$ as answer and adds
  $(ID,x,D_{ID},P_{ID},c')$ to \textbf{K}; otherwise, $\mathcal {C}$
  aborts.
\end{enumerate}

\medskip
\noindent \emph{Ring-Sign Queries}: $\mathcal {A}_I$ chooses a group
of $n$ users whose identities form the set
$L_{ID}=\{ID_1,...,ID_n\}$ and their corresponding public keys form
the set $L_{PK}=\{P_1,...,P_n\}$, and may ask a ring signature on a
message $M$ of this group. On receiving a {\sf Ring-Sign} query
$RS(M,L_{ID},L_{PK})$, $\mathcal {C}$ creates a ring signature as
follows:
\begin{enumerate}
  \item Choose a random index $s\in\{1,...,n\}$.
  \item For all $i\in\{1,...,n\}\backslash\{s\}$, choose $r_i\in Z_q^*$ uniformly at
  random,
  compute $y_i=g^{r_i}$.
  \item For all $i\in\{1,...,n\}\backslash\{s\}$, compute
  $h_i=H_2(M||L_{ID}||L_{PK}||y_i)$.
  \item Choose $h_s\in Z_q^*, V\in G_1$ at random.
  \item Compute $y_s=e(V-(\sum_{i\neq s}r_i)P,P)e(\sum_{i=1}^nh_iQ_i,-P_0)e(\sum_{i=1}^nh_iP_i,-U)$
  (Where $U=H_3(M||$ $L_{ID}||L_{PK}),$ $Q_i=H_1(ID_i)$). If $y_s=1_{G_2}$ or $y_s = y_i$ for some $i\neq s$, then goto step 4.
  \item Set $H_2(M||L_{ID}||L_{PK}||y_s)=h_s$.
  \item Return $(M,L_{ID},L_{PK},\sigma=\{(y_1,...,y_n),V\})$ as answer.
\end{enumerate}

\noindent \emph{Forgery:} Finally, $\mathcal {A}_I$ outputs a tuple
$(M^*,L_{ID}^*=\{ID_1^*,...,ID_n^*\},L_{PK}^*=\{P_1^*,...,P_n^*\},\sigma^*=\{(y_1^*,...,y_n^*),V^*\})$
which means  $\sigma^*$ is a ring signature on message $M^*$ on
behalf of the group specified by identities in $L_{ID}^*$ and the
corresponding public keys in $L_{PK}^*$. It is required that
$\mathcal {C}$ does not know the private key of any member in this
group, $L_{ID}^* \cap ((L_1\cap L_2)\cup L_3)= \emptyset$ and the
ring signature $\sigma^*$ on message $M^*$ on behalf of the group
must be valid (\textbf{Event 1}). Now, applying the `ring forking
lemma' \cite{Herranz}, if $\mathcal {A}_I$ succeeds in outputting a
valid ring signature $\sigma^*$ with probability $\epsilon\geq
7P_n^{q_{H_1}}/2^\ell$ in a time span $t$ in the above interaction,
then within time $2t$ and probability
$\geq\epsilon^2/66P_n^{q_{H_1}}$, $\mathcal {C}$ can get two valid
ring signatures
$(M^*,L_{ID}^*,L_{PK}^*,\sigma^*=\{(y_1^*,...,y_n^*),V^*\})$ and
$(M^*,L_{ID}^*,L_{PK}^*,\sigma'^*=\{(y_1^*,...,y_n^*),V'^*\})$. From
these two valid ring signatures, $\mathcal {C}$ obtains
$$e(V^*,P)=y_1^*\cdot ...\cdot y_n^*e(\sum_{i=1}^nh_i^*P_i^*,U^*)e(\sum_{i=1}^nh_i^*Q_i^*,P_0)$$
and
$$e(V'^*,P)=y_1^*\cdot ...\cdot y_n^*e(\sum_{i=1}^nh_i'^*P_i^*,U^*)e(\sum_{i=1}^nh_i'^*Q_i^*,P_0)$$
Where $U^*=H_3(M^*||L_{ID}^*||L_{PK}^*)$, $Q_i^*=H_1(ID_i^*)$,
$h_i^*=H_2(M^*,L_{ID}^*,L_{PK}^*,y_i^*)$, $h_i'^*=H_2'(M^*,$
$L_{ID}^*,L_{PK}^*,y_i^*)$, and for some $s\in\{1,...,n\}$,
$h_s^*\neq h_s'^*$, while for $i\in\{1,...,n\}\backslash\{s\}$,
$h_i^*=h_i'^*$. From the above two equations we have
$$e(V^*-V'^*,P)=e((h_s^*-h_s'^*)P_s^*,U^*)e((h_s^*-h_s'^*)Q_s^*,P_0)$$
\indent At this stage, $\mathcal {C}$ may find the item
$(M^*,L_{ID}^*,L_{PK}^*,\beta^*,U^*,c''^*)$ from $\bf H_3$,
$(ID_s^*,\alpha_s^*,Q_s^*,c_s^*)$ from $\bf H_1$,
$(ID_s^*,x_s^*,D_s^*,P_s^*,{c'}_s^*)$ from $\bf K$. There are three
cases in which $\mathcal {C}$ can successfully solve the CDH
problem.
\begin{itemize}
  \item \textbf{Case 1:} $c_s^*=1,c''^*=0$, this means
  $Q_s^*=\alpha_s^*bP,U^*=\beta^*P$. In this case,
  $e(V^*-V'^*,P)=e((h_s^*-h_s'^*)(\beta^*P_s^*+\alpha_s^*abP),P)$. So, $\mathcal {C}$ can get
$abP={\alpha_s^*}^{-1}((h-h')^{-1}(V^*-V'^*)-\beta^*P_s^*)$.
  \item \textbf{Case 2:} ${c'}_s^*=1,c''^*=1,c_s^*=0,x_s^*\neq\bot$, and $P_s^*=x_s^*aP,U^*=\beta^*bP,
  Q_s^*=\alpha_s^*P$, $\mathcal {C}$ can get
$abP=(x_s^*\beta^*)^{-1}((h-h')^{-1}(V^*-V'^*)-\alpha_s^*P_0)$.
  \item \textbf{Case 3:} ${c'}_s^*=1,c''^*=1,c_s^*=1,x_s^*\neq\bot$, and $P_s^*=x_s^*aP,U^*=\beta^*bP,
  Q_s^*=\alpha_s^*bP$, $\mathcal {C}$ can get
$abP=(x_s^*\beta^*+\alpha_s^*)^{-1}(h-h')^{-1}(V^*-V'^*)$ (Note the
probability that $x_s^*\beta^*+\alpha_s^*=0$ is negligible).
\end{itemize}

\medskip
\noindent \emph{Probability of Success}: Now we determine the value
of $\delta$ and consider the probability for $\mathcal {C}$ to
successfully solve the given CDH problem. The probability that
$\mathcal {C}$ does not abort in all the $q_K$
\emph{Partial-Private-Key Queries} and $q_{Pr}$ \emph{Private-Key
Queries} is at least $\delta^{q_K+q_{Pr}}$. The probability that the
forged ring signature is helpful for $\mathcal {C}$ to solve the CDH
problem is $Pr[\bf (Case 1\vee Case 2\vee Case 3)\wedge
\textbf{Event 1}]$$\leq (1-\delta)^n$. So the combined probability
is $\delta^{q_K+q_{Pr}}(1-\delta)^n$. We can find the value of
$\delta$ that maximize this probability is
$\frac{q_K+q_{Pr}}{q_K+q_{Pr}+n}$ and the maximized probability is
$(\frac{q_K+q_{Pr}}{q_K+q_{Pr}+n})^{q_K+q_{Pr}}(\frac{n}{q_K+q_{Pr}+n})^n$.\\
\indent Based on the bound from the ring forking lemma
\cite{Herranz}, if $\mathcal {A}_I$ succeeds in time $\leq t$ with
probability $\epsilon\geq 7P_n^{q_{H_1}}/2^\ell$, then the CDH
problem in $G_1$ can be solved by $\mathcal {C}$ within time
$2(t+q_{H_1}T_1+q_{H_2}T_2+q_{H_3}T_3+q_KT_K+q_PT_P+q_{Pr}T_{Pr}+q_ST_S)$
and with probability $\geq
((\frac{q_K+q_{Pr}}{q_K+q_{Pr}+n})^{q_K+q_{Pr}}\cdot$
$(\frac{n}{q_K+q_{Pr}+n})^n\epsilon)^2/66P_n^{q_{H_1}}$.

%%%%%%%%%%%%%%%%%%%%%%%%%%%%%%%%%%%%%%%%%%%%%%%%%%%%%%%%%
%                                                      %
%  Proof of Theorem 2                                  %
%                                                      %
%%%%%%%%%%%%%%%%%%%%%%%%%%%%%%%%%%%%%%%%%%%%%%%%%%%%%%%%%

\section{Proof of Theorem~\ref{Theorem-2}}\label{Appendix-Proof-Th2}
\emph{Proof}. Let $\mathcal {A}_{II}$ be our type II adversary,
$\mathcal {C}$ be a CDH attacker who receives a random instance
$(P,aP,bP)$ and has to compute the value of $abP$.

\medskip
\noindent \emph{Setup:} $\mathcal {C}$
 generates the KGC's {\sf masterkey} $\kappa\in Z_q^*$ and
 system parameters {\sf param}=$(G_1,G_2,e, P,g,
 P_0,$ $H_1,H_2,H_3)$. When the simulation is started, $\mathcal {A}_{II}$ is
provided with {\sf param} and the {\sf masterkey} $\kappa$.

\medskip
\noindent \emph{Training:} $\mathcal {A}_{II}$ can ask $\mathcal
{C}$ $H_1,H_2,H_3$, Public-Key, Private-Key, and Ring-Sign queries.
Since $\mathcal {A}_{II}$ has access to the {\sf masterkey}
$\kappa$, he can do {\sf Partial-Private-Key-Extract} himself.
$\mathcal {C}$ also maintains four lists, namely ${\bf H_1}$
contains items of the form $(ID,Q_{ID})$, ${\bf H_2}$ contains items
of the form $(M,L_{ID},L_{PK},y,h)$, ${\bf H_3}$ contains items of
the form $(M,L_{ID},L_{PK},\beta,U)$ and \textbf{K} contains items
of the form $(ID,x,P_{ID})$ to store the answers used. All of these
four lists are initially empty.  $\mathcal{C}$ also maintains two
lists $L_1,L_2$, the function of these two lists are the same as
mentioned in Game 2 Section~\ref{Sub-Adver-Model}.

\medskip
\noindent $H_1$ \emph{Queries}: On receiving a query  $H_1(ID)$. If
$(ID,Q_{ID})$ exists in $\bf H_1$ then $\mathcal {C}$ returns
$Q_{ID}$ as answer. Otherwise, $\mathcal {C}$ picks a random
$Q_{ID}\in G_1^*$ which has not been used in the former $H_1$
\emph{Queries}, then returns $Q_{ID}$ as answer and adds
$(ID,Q_{ID})$ to $\bf H_1$.

\medskip
\noindent $H_2$ \emph{Queries}: On receiving a query
$H_2(M||L_{ID}||L_{PK}||y)$, $\mathcal {C}$ first checks whether
there exists an item $(M,L_{ID},L_{PK},y,h)$ in $\bf H_2$, if so,
returns $h$ as answer. Otherwise, $\mathcal {C}$ picks a random
$h\in Z_q^*$ which has not been used in the former $H_2$
\emph{Queries}, then returns $h$ as answer and adds
$(M,L_{ID},L_{PK},y,h)$ to $\bf H_2$.

\medskip
\noindent $H_3$ \emph{Queries}: Whenever receives a query
$H_3(M||L_{ID}||L_{PK})$, $\mathcal {C}$ first checks whether there
exists an item $(M,L_{ID},L_{PK},\beta,U)$ in $\bf H_3$, if so,
returns $U$ as answer. Otherwise, $\mathcal {C}$ picks a random
$\beta \in Z_q^*$ which has not been used in the former $H_3$
\emph{Queries}, computes $U=\beta aP$, then adds
$(M,L_{ID},L_{PK},\beta ,U)$ to $\bf H_3$ and returns $U$ as answer.

\medskip
\noindent \emph{Public-Key Queries}: On receiving a query $PK(ID)$
\begin{enumerate}
  \item If there is an item $(ID,x,P_{ID},c)$ exists in \textbf{K}, then $\mathcal {C}$ returns $P_{ID}$ as
  answer.
  \item Otherwise, $\mathcal {C}$ first  flips a coin $c\in \{0,1\}$
    that yields 0 with probability $\delta$ and 1 with
    probability $1-\delta$, then picks a random $x\in Z_q^*$. If
    $c=0$, $\mathcal {C}$ sets $P_{ID}=xP$, returns $P_{ID}$ as answer
    and adds $(ID,x,P_{ID},c)$ to \textbf{K}.
    Else $c=1$, $\mathcal {C}$ sets $P_{ID}=xbP$, returns $P_{ID}$ as answer and adds $(ID,x,P_{ID},c)$ to
    \textbf{K}.
\end{enumerate}

\noindent \emph{Public-Key-Replacement Queries}: On receiving a
query $PKR(ID,P'_{ID})$ ($\mathcal {C}$ sets $L_1=L_1\cup\{ID\}$),
$\mathcal {C}$ first makes a $PPK (ID)$ query to obtain an item
$(ID,x,P_{ID},c)$, then sets $x=\bot$,
 $P_{ID}=P'_{ID}$, and updates the item $(ID,x,P_{ID},c)$ in \textbf{K} to record this replacement.

\medskip \noindent \emph{Private-Key Queries}: On receiving a query
$PrK(ID)$, if $ID\in L_1$ $\mathcal {C}$ returns $\bot$, otherwise
\begin{enumerate}
  \item If there is an item $(ID,x,P_{ID},c)$ in \textbf{K}, when $c=0$
      $\mathcal {C}$ sets $L_2=L_2\cup\{ID\}$, returns $(x,D_{ID})$ as answer
      (where $D_{ID}=\kappa H_1(ID)$ is the partial private key of the user whose identity is $ID$);
      while $c=1$, $\mathcal {C}$ aborts.
  \item Otherwise, $\mathcal {C}$ first makes a $PK(ID)$ query to obtain an item
  $(ID,x,P_{ID},c)$; when $c=0$, $\mathcal {C}$ sets $L_2=L_2\cup\{ID\}$, returns $(x,D_{ID})$ as answer;
       while $c=1$, $\mathcal {C}$ aborts.
\end{enumerate}

\medskip
\noindent \emph{Ring-Sign Queries}: $\mathcal {A}_{II}$ chooses a
group of $n$ users whose identities form the set
$L_{ID}=\{ID_1,...,ID_n\}$ and their corresponding public keys form
the set $L_{PK}=\{P_1,...,P_n\}$. On receiving a {\sf Ring-Sign}
query $RS(M,L_{ID},L_{PK})$, $\mathcal {C}$ creates a ring signature
as follows:
\begin{enumerate}
  \item Choose a random index $s\in\{1,...,n\}$;
  \item For all $i\in\{1,...,n\}\backslash\{s\}$, choose $r_i\in Z_q^*$ uniformly at random,
  compute $y_i=g^{r_i}$.
  \item For all $i\in\{1,...,n\}\backslash\{s\}$, set
  $h_i=H_2(M||L_{ID}||L_{PK}||y_i)$.
  \item Randomly choose $h_s\in Z_q^*, V\in G_1$.
  \item Compute $y_s=e(V-(\sum_{i\neq s}r_i)P,P)e(\sum_{i=1}^nh_iQ_i,-P_0)e(\sum_{i=1}^nh_iP_i,-U)$
  (Where $U=H_3(M||$ $L_{ID}||L_{PK}),$ $Q_i=H_1(ID_i)$). If $y_s=1_{G_2}$ or $y_s = y_i$ for some $i\neq s$, then goto step 4.
  \item Set $H_2(M||L_{ID}||L_{PK}||y_s)=h_s$.
  \item Return $(M,L_{ID},L_{PK},\sigma=\{(y_1,...,y_n),V\})$ as answer.
\end{enumerate}

\noindent \emph{Forgery:} Finally, $\mathcal {A}_{II}$ outputs a
tuple
$(M^*,L_{ID}^*=\{ID_1^*,...,ID_n^*\},L_{PK}^*=\{P_1^*,...,P_n^*\},\sigma^*=\{(y_1^*,...,y_n^*),V^*\})$
which implies that $\sigma^*$ is a ring signature on message $M^*$
on behalf of the group specified by identities in $L_{ID}^*$ and the
corresponding public keys in $L_{PK}^*$. It is required that
$(M^*,\sigma^*)$ is a valid message and ring signature pair,
$L_{ID}^* \cap (L_1\cup L_2)= \emptyset$ and $\mathcal {C}$ does not
know the private key of any member in the group specified by
$L_{ID}^*$ and $L_{PK}^*$ (\textbf{Event 1}). Now, Applying the
`ring forking lemma' \cite{Herranz}, $\mathcal {C}$ gets two valid
ring signatures
$(M^*,L_{ID}^*,L_{PK}^*,\sigma^*=\{(y_1^*,...,y_n^*),V^*\})$ and
$(M^*,L_{ID}^*,L_{PK}^*,\sigma'^*=\{(y_1^*,...,y_n^*),V'^*\})$. From
these two ring signatures, $\mathcal {C}$ obtains
$$e(V^*,P)=y_1^*\cdot ...\cdot y_n^*e(\sum_{i=1}^nh_i^*P_i^*,U^*)e(\sum_{i=1}^nh_i^*Q_i^*,P_0)$$
and
$$e(V'^*,P)=y_1^*\cdot ...\cdot y_n^*e(\sum_{i=1}^nh_i'^*P_i^*,U^*)e(\sum_{i=1}^nh_i'^*Q_i^*,P_0)$$
\noindent Where $U^*=H_3(M^*||L_{ID}^*||L_{PK}^*)$,
$Q_i^*=H(ID_i^*)$, $h_i^*=H_2(M^*,L_{ID}^*,L_{PK}^*,y_i^*)$ and
$h_i'^*=H_2'(M^*,L_{ID}^*,L_{PK}^*,y_i^*)$. The hash functions $H_2$
and $H_2'$ satisfy: for some $s \in\{1,...,n\}$, $h_s^*\neq h_s'^*$,
while $i\in\{1,...,n\}\backslash\{s\}$, $h_i^*=h_i'^*$. From the
above two equations we have
$$e(V^*-V'^*,P)=e((h_s^*-h_s'^*)P_s^*,U^*)e((h_s^*-h_s'^*)Q_s^*,P_0)$$
\indent At this point, $\mathcal {C}$ may find the item
$(M^*,L_{ID}^*,L_{PK}^*,\beta^* ,U^*)$ from $\bf H_3$,
$(ID_s^*,Q_s^*)$ from $\bf H_1$ and $(ID_s^*,x_s^*,P_s^*,c^*)$ from
$\bf K$. Since $U^*=\beta^* aP,P_s^*=x_s^*bP$, $\mathcal {C}$ has
the following
$$e(V^*-V'^*,P)=e((h_s^*-h_s'^*)(x_s^*\beta^*abP+\kappa Q_s^*),P)$$
This implies
\begin{eqnarray*}
V^*-V'^*&=&(h_s^*-h_s'^*)(x_s^*\beta^*abP+\kappa
Q_s^*)\\
\end{eqnarray*}
Hence, $\mathcal {C}$ can obtain
$abP=(x_s^*\beta^*)^{-1}((h_s^*-h_s'^*)^{-1}(V^*-V'^*)-\kappa
Q_s^*)$.

\medskip
\noindent \emph{Probability of Success}: Now we determine the value
of $\delta$ and consider the probability for $\mathcal {C}$ to
successfully solve the given CDH problem. The probability that
$\mathcal {C}$ does not abort in all the $q_K$ \emph{Private-Key
Queries} is $\delta^{q_K}$. The probability that $\mathcal {A}_{II}$
forged a valid ring signature which $\mathcal {C}$ does not know any
private key of the group members' involved in the ring signature
 is $(1-\delta)^n$. So the combined
probability ($Pr$[\textbf{Event 1}]) is $\delta^{q_K}(1-\delta)^n$.
We can find the value of $\delta$ that maximize this probability is
$\frac{q_K}{q_K+n}$ and the maximized probability is
$(\frac{q_K}{q_K+n})^{q_K}(\frac{n}{q_K+n})^n$.\\
\indent Based on the bound from the ring forking lemma
\cite{Herranz}, if $\mathcal {A}_{II}$ succeeds in time $\leq t$
with probability $\epsilon\geq 7P_n^{q_{H_1}}/2^\ell$, then the CDH
problem in $G_1$ can be solved by $\mathcal {C}$ within time
$2(t+q_{H_1}T_1+q_{H_2}T_2+q_{H_3}T_3+q_KT_{Pr}+q_PT_P+q_ST_S)$ and
with
probability $\geq ((\frac{q_K}{q_K+n})^{q_K}(\frac{n}{q_K+n})^n\epsilon)^2/66P_n^{q_{H_1}}$.\\


\begin{thebibliography}{99}
\bibitem{M. Abe} M. Abe, M. Ohkubo, and K. Suzuki. 1-out-of-n signatures from a variety of keys.
ASIACRYPT 2002, \emph{Lecture Notes in Computer Science}, vol. 2501,
pages 415-432, Springer-Verlag, 2002.

\bibitem{Al-Riyami} S. Al-Riyami and K. Paterson. Certificateless public key cryptography. Asiacrypt 2003,
\emph{Lecture Notes in Computer Science}, vol. 2894, pages 452-473,
Springer-Verlag, 2003.

%\bibitem{Al-Riyami2} S. Al-Riyami and K. Paterson. ``CBE from CL-PKE: a generic construction
%and efficient schemes''. PKC 2005, \emph{Lecture Notes in Computer
%Science}, vol. 3386, pages 398-415, Springer-Verlag, 2005.

%\bibitem{M. Au} M. Au, J. Liu, T. Yuen and D. Wong. ID-Based Ring Signature Scheme Secure in the Standard
%Model. Advances in Information and Computer Security, \emph{Lecture
%Notes in Computer Science}, vol. 4266, pages 1-16, Springer-Verlag,
%2006.

%\bibitem{Joonsang Baek} J. Baek, R. Safavi-Naini, and W. Susilo.
%Certificateless public key encryption without pairing. ISC 2005,
%\emph{Lecture Notes in Computer Science}, vol. 3650, pages 134-148,
%Springer-Verlag, 2005.

\bibitem{Bellare} M. Bellare and P. Rogaway. Random oracles are
practical: A paradigm for designing efficient protocols. ACM CCS
1993, pages 62-73, 1993.

%\bibitem{BLS} D. Boneh, B. Lynn, and H. Shacham. Short signatures from the weil
%pairing. Asiacrypt 2001, \emph{Lecture Notes in Computer Science},
%vol. 2248, pages 514-532, Springer-Verlag, 2001.

\bibitem{Dan Boneh} D. Boneh, C. Gentry, B. Lynn and H. Shacham.
Aggregate and verifiably encrypted signatures from bilinear maps.
EUROCRPYT 2003, \emph{Lecture Notes in Computer Science}, vol. 2656, pp. 416-432,
Springer-Verlag, 2003.

\bibitem{Bresson} E. Bresson, J. Stern, and M. Szydlo. Threshold
ring signature and applications to ad-hoc groups. Crypto 2002,
\emph{Lecture Notes in Computer Science}, vol. 2442, pages 465-480,
Springer-Verlag, 2002.

%\bibitem{Chan} T. Chan, K. Fung, J. Liu and V. Wei. Blind spontaneous
%anonymous groupsignatures for ad hoc groups. Proceedings of ESAS'04,
%\emph{Lecture Notes in Computer Science}, vol. 3313, pages 82-84
%Springer-Verlag, 2005.

    \bibitem{CZQWZ09}	W. Chen, L. Zhang, B. Qin, Q. Wu, H. Zhang. Certificateless One-Way Authenticated Two-Party Key Agreement Protocol, Fifth International Conference on Information Assurance and Security (IAS 09), IEEE, pp. 483-486, 2009.
\bibitem{CZQWZ10}	L. Chen, L. Zhang, B. Qin, Q. Wu, H. Zhang. Cryptanalysis of a Certificateless Encryption Scheme, 2010 International Conference on Computer Design and Applications (ICCDA 2010), IEEE, pp. V5-536 - V5-539, 2010.


%\bibitem{C. Chunjie} C. Chunjie, M. Jianfeng
%M. Sangjae. Provable efficient certificateless group key exchange
%Protocol. \emph{Wuhan University Journal of Natural Sciences}, vol.
%12 pages 041-045 No.1 2007.

%\bibitem{Sherman} S. Chow, C. Boyd, and J. Nieto. Security-mediated certificateless
%cryptography. PKC 2006, \emph{Lecture Notes in Computer Science},
%vol. 3958, pages 508-524, Springer-Verlag, 2006.
%
%\bibitem{S.M. Chow} S. Chow , S. Yiu  and L. Hui. Efficient identity based ring
%signature. ACNS 2005, \emph{Lecture Notes in Computer Science}, vol.
%3531, page 499, Springer-Verlag, 2005.


%\bibitem{S.M. Chow2} S. Chow, R. Lui, L. Hui, and S. Yiu.
%Identity based ring signature: why, how and what next. EuroPKI 2005,
%\emph{Lecture Notes in Computer Science}, vol. 3545, pp. 144-161, Springer-Verlag, 2005.

\bibitem{S. Chow3} S. Chow and W. Yap. Certificateless Ring
Signatures. Cryptology ePrint Archive, Report 2007/236.

%\bibitem{chow 4} S. Chow and D. Wong. Anonymous Identification and
%Designated-Verifiers Signatures from Insecure Batch Verification.
%EuroPKI 2007, \emph{LNCS} 4582, pp. 203-219, Springer-Verlag, 2007.

\bibitem{DLZZS17} F. Dai, M. Luo, Y. Zhang, L. Zhang and Y. Sun. A Fault-Tolerant Batch Verification Scheme for Cloud Assisted VANETs, 2nd International Conference on Applied Mechanics, Electronics and Mechatronics Engineering (AMEME 2017), pages 337-342, 2017.

\bibitem{Y. Dodis} Y. Dodis, A. Kiayias, A. Nicolosi, and V. Shoup. Anonymous
identification in ad hoc groups. EUROCRYPT 2004, Springer-Verlag,
\emph{Lecture Notes in Computer Science}, vol. 3027, pages 609-626,
Springer-Verlag, 2004.

\bibitem{DZL14}	Z. Dong, L. Zhang, J. Li, Security Enhanced Anonymous Remote User Authentication and Key Agreement for Cloud Computing, 17th International Conference on Computational Science and Engineering (CSE 2014), IEEE, pp. 1746-1751, 2014.


%\bibitem{Gorantla} M. Gorantla, A. Saxena. An efficient certificateless signature
%scheme. CIS 2005, \emph{Lecture Notes in Computer Science}, vol.
%3802 pages 110-116, Springer-Verlag, 2005.

\bibitem{Herranz F} J. Herranz and G. Saez. Forking lemmas for ring signature schemes. INDOCRYPT 2003,
\emph{Lecture Notes in Computer Science}, vol. 2904, pp. 266-279, Springer-Verlag, 2003.

\bibitem{Herranz} J. Herranz and G. Saez. New identity-based ring signature
schemes. ICICS 2004, \emph{Lecture Notes in Computer Science}, vol.
3269, pages 27-39, Springer-Verlag, 2004.

\bibitem{Hu} B. Hu, D. Wong, Z. Zhang and X. Deng. Key replacement attack
against a generic construction of certificateless signature. ACISP
2006, \emph{Lecture Notes in Computer Science}, vol. 4058, pages
235-346, Springer-Verlag, 2006.

%\bibitem{ACISP 2007}X. Huang, Y. Mu, W. Susilo,
%D. Wong and W. Wu. Certificateless signature revisited. ACISP 2007,
%\emph{Lecture Notes in Computer Science}, vol. 4586, pages 308-322,
%Springer-Verlag, 2007.
%
\bibitem{Huang} X. Huang, W. Susilo, Y. Mu and F. Zhang. On the security of a certificateless
signature scheme. CANS 2005, \emph{Lecture Notes in Computer
Science}, vol. 3810, pages 13-25, Springer-Verlag, 2005.

%\bibitem{Huang07} X. Huang, W. Susilo, Y. Mu and F. Zhang. Certificateless designated verifier signature
%schemes. AINA, Proceedings-20th International Conference on Advanced
%Information Networking and Applications, vol. 2, pages 15-19, 2006.

\bibitem{X. Li} X. Li, K. Chen and L. Sun. Certificateless signature and proxy signature schemes
from bilinear pairings. \emph{Lithuanian Mathematical Journal}, vol.
45, pages 76-83, Springer-Verlag, 2005.

\bibitem{LZ13} B. Liu, L. Zhang. An Improved Identity-based Batch Verification Scheme for VANETs, 5th International Conference on Intelligent Networking and Collaborative Systems (INCos 2013), IEEE, pp. 809-814, 2013.

%\bibitem{J. Liu AS} J. Liu, M. Au and W. Susilo. Self-Generated-Certificate Public
%Key Cryptography and Certificateless Signature/Encryption Scheme in
%the Standard Model. 2007 ACM Symposium on InformAtion, Computer and
%Communications Security (ASIACCS'07), 2007.

%\bibitem{J. Liu ww} J. Liu, V. Wei, and D. Wong. Linkable spontaneous anonymous
%group signature for ad hoc groups. ACISP'04, \emph{Lecture Notes in
%Computer Science}, vol. 3108, pages 325¨C335, Springer-Verlag, 2004.

\bibitem{J. Liu} J. Liu and D. Wong. On the security models of (threshold) ring signature
schemes. ICISC 2004, \emph{Lecture Notes in Computer Science}, vol.
3506, pages 204-217, Springer-Verlag, 2005.

%\bibitem{B. Libert} B. Libert and J. Quisquater. On constructing certificateless cryptosystems
%from identity based encryption. PKC 2006, \emph{Lecture Notes in
%Computer Science}, vol. 3958, pages 474-490, Springer-Verlag, 2006.

%\bibitem{Ma. Chunbo} C. Ma, F. Ao and D. He. Certificateless group inside
%signature. ISADS 2005, Proceedings-2005 International Symposium on
%Autonomous Decentralized Systems, vol. 2005, pages 194-200, 2005.

\bibitem{MZZ10a}S. Miao, F. Zhang, L. Zhang. On the Security of a Certificateless Signature Scheme, 2nd International Conference on Signal Processing Systems (ICSPS 2010), IEEE, pp. V2-457 - V2-461, 2010.
    
\bibitem{MZZ10b}S. Miao, F. Zhang, L. Zhang. Cryptanalysis of a Certificateless Multi-receiver Signcryption Scheme, 2010 International Conference on Multimedia Information Networking and Security (MINES 2010), IEEE, pp. 593-597, 2010.

\bibitem{L. Nguyen} L. Nguyen. Accumulators from bilinear pairings and
applications. CT-RSA 2005, \emph{Lecture Notes in Computer Science},
vol. 3376, pages 275-292, Springer-Verlag, 2005.

%\bibitem{Pointcheval} D. Pointcheval and J. Stern. Security proofs for signature schemes.
%Eurocrypt 1996, \emph{Lecture Notes in Computer Science},
%vol. 1070, pages 387-398, Springer-Verlag, 1996.

\bibitem{R. Rivest} R. Rivest, A. Shamir and Y. Tauman. How to leak a secret. Asiacrypt'01,
\emph{Lecture Notes in Computer Science}, vol. 2248, pages 552-565,
Springer-Verlag, 2001.

\bibitem{Shamir} A. Shamir. Identity based cryptosystems and signature schemes.
Crypto'84, \emph{Lecture Notes in Computer Science}, vol. 196, pages
47-53, Springer-Verlag, 1984.

%\bibitem{Licheng Wang} L. Wang, Z. Cao, X. Li, and H. Qian. Certificateless threshold signature
%schemes. CIS 2005, \emph{Lecture Notes in Computer Science}, vol.
%3802, pages 104-109, Springer-Verlag, 2005.

%\bibitem{S. Wang} S. Wang, Z. Cao, X. Dong. Certificateless authenticated key agreement based on the MTI/CO
%protocol. \emph{Journal of Information and Computational Science},
%vol. 3, pages 575-581, 2006.

%\bibitem{Duncan} D. Wong, K. Fung, J. Liu, and V. Wei. On
%the RSCode construction of ring signature schemes and a threshold
%setting of RST. ICICS 2004, \emph{Lecture Notes in Computer
%Science}, vol. 2836, pages 34-46, 2004.

\bibitem{Yap} W. Yap, S. Heng, and B. Goi. An efficient certificateless signature
scheme. EUC Workshops 2006, \emph{Lecture Notes in Computer
Science}, vol. 4097, pages 322-331, Springer-Verlag, 2006.

%\bibitem{S. Yijuan} S. Yijuan, L. Jianhua. Two-party authenticated key
%agreement in certificateless public key cryptography. \emph{Wuhan
%University Journal of Natural Sciences}, vol. 12 pages 071-074 No.1
%2007.

%\bibitem{Yum E} D. Yum and P. Lee. Generic construction of certificateless
%encryption. ICCSA'04, \emph{Lecture Notes in Computer Science}, vol.
%3043, pages 802-811, Springer-Verlag, 2004.

\bibitem{YZHMSZ10}	H. Yuan, F. Zhang, X. Huang, Y. Mu, W. Susilo, L. Zhang. Certificateless Threshold Signature Scheme from Bilinear Pairings, Information Sciences, 180(23), 4714-4728, 2010.

\bibitem{Yum} D. Yum and P. Lee. Generic construction of certificateless signature. ACISP 2004, \emph{Lecture Notes in Computer
Science}, vol. 3108, pages 200-211, Springer-Verlag, 2004.

\bibitem{Zhang Kim} F. Zhang and K. Kim. ID-Based blind signature and ring
signature from pairings. ASIACRYPT 2002, \emph{Lecture Notes in
Computer Science volume}, vol. 2501, pages 533-547, Springer-Verlag,
2002.

%\bibitem{F. Zhang} F. Zhang, R. Safavi-Naini, W. Susilo. An efficient signature scheme from bilinear pairings and its
%applications. Public Key Cryptography-PKC 2004, \emph{Lecture Notes
%in Computer Science}, vol. 2947, pages 277-290, Springer-Verlag,
%2004.
\bibitem{agg1}	L. Zhang, F. Zhang. Security Model for Certificateless Aggregate Signature Schemes. 2008 International Conference on Computational Intelligence and Security (CIS 2008), IEEE, pp. 364-368, 2008.
\bibitem{agg2}	L. Zhang, B. Qin, Q. Wu, F. Zhang. Novel Efficient Certificateless Aggregate Signatures, The 18th Symposium on Applied algebra, Algebraic algorithms, and Error Correcting Codes (AAECC 2009), \emph{Lecture Notes in
Computer Science volume}, vol. 5527, pp. 235-238, Springer-Verlag, 2009.

\bibitem{agg3}	L. Zhang, F. Zhang. A New Certificateless Aggregate Signature Scheme, Computer Communications, 32(6), 1079-1085, 2009.
\bibitem{agg4}	L. Zhang, B. Qin, Q. Wu, F. Zhang. Efficient Many-to-One Authentication with Certificateless Aggregate Signatures, Computer Networks, 54(14), 2482-2491, 2010.
%----------------------------- agg









%\bibitem{ZWQDZL13}	L. Zhang, Q. Wu, B. Qin, J. Domingo-Ferrer, P. Zeng, J. Liu, Ruiying Du, A Generic Construction of Proxy Signatures from Certificateless Signatures, The 27th IEEE International Conference on Advanced Information Networking and Applications (AINA 2013), IEEE, pp. 259-266, 2013.
%
%\bibitem{LWQDLS14}	Lei Zhang, Qianhong Wu, Bo Qin, Hua Deng, Jianwei Liu, Wenchang Shi, Provably Secure Certificateless Authenticated Asymmetric Group Key Agreement, The 10th International Conference on Information Security Practice and Experience (ISPEC 2014), Lecture Notes in Computer Science, vol. 8434, pp. 496¨C510, 2014.
    


%\bibitem{Zhang Kim}	Lei Zhang*, Futai Zhang. Certificateless Signature and Blind Signature, µç×Ó¿Æѧѧ¿¯£¨Ó¢Îİ棩, 25(5), 629-635, 2008.

%\bibitem{ZZWD10}	L. Zhang, F. Zhang, Qianhong Wu, Josep Domingo-Ferrer. Simulatable Certificateless Two-Party Authenticated Key Agreement Protocol, Information Sciences, 180(6), 1020¨C1030, 2010.

%\bibitem{Zhang Kim}	Lei Zhang*, Futai Zhang, Xinyi Huang. A Secure and Efficient Certificateless Signature Scheme Using Bilinear Pairing, µç×Óѧ±¨Ó¢ÎÄ°æ, 18(1), 145-148, 2009.



%\bibitem{Zhang Kim}	Lei Zhang, Futai Zhang, Bo Qin, Shubo Liu. Provably-Secure Electronic Cash Based on Certificateless Partially-Blind Signatures, Electronic Commerce Research and Applications, 10(5), 545-552, 2011.
%
%\bibitem{Zhang Kim}	Lei Zhang*, Futai Zhang, Qianhong Wu, Delegation of Signing Rights using Certificateless Proxy Signatures, Information Sciences, 184(1), 298-309, 2012.
%
%\bibitem{Zhang Kim}	L. Zhang, Certificateless One-Pass and Two-Party Authenticated Key Agreement Protocol and Its Extensions, Information Sciences, 293(2015), 182-195, 2015.
%\bibitem{Zhang Kim}	Lei Zhang*, Qianhong Wu, Josep Domingo-Ferrer, Bo Qin, Peng Zeng, Signatures in Hierarchical Certificateless Cryptography: Efficient Constructions and Provable Security,   Information Sciences, 272 (2014), 223¨C237, 2014.
%\bibitem{Zhang Kim}	Lei Zhang, Qianhong Wu, Bo Qin, Hua Deng, Jiangtao Li, Jianwei Liu, Wenchang Shi, Certicateless and Identity-based Authenticated Asymmetric Group Key Agreement, International Journal of Information Security, 16(5), 559¨C576, October 2017.
%
%
%
%    ---------------------------------

\bibitem{ZWDQ10}L. Zhang,¡¡Q. Wu,¡¡J. Domingo-Ferrer,¡¡B. Qin. Hierarchical Certificateless Signatures, 2010 IEEE/IFIP 8th International Conference on Embedded and Ubiquitous Computing (EUC), IEEE, pp. 572-577, 2010.

\bibitem{ZZ08}	L. Zhang, F. Zhang. A New Provably Secure Certificateless Signature Scheme, 2008 IEEE International Conference on Communications (ICC 2008), pp. 1685-1689, IEEE, 2008.

\bibitem{ZZW07} L. Zhang, F. Zhang, W. Wu. A Provably Secure Ring Signature
Scheme in Certificateless Cryptography. ProvSec 2007, \emph{Lecture
Notes in Computer Science}, vol. 4784, pages 103-121,
Springer-Verlag, 2007.

\bibitem{Z. Zhang} Z. Zhang, D. Wong, J. Xu and D. Feng. Certificateless public-key
signature: security model and efficient construction. ACNS 2006,
\emph{Lecture Notes in Computer Science}, vol. 3989, pages 293-308,
Springer-Verlag, 2006.
\end{thebibliography}
\end {document}